\documentclass[conference]{IEEEtran}
\usepackage[mathcal]{euscript}
\usepackage{bbm,dsfont,mathrsfs,fixmath}
\usepackage{stmaryrd}
\usepackage{amsmath,amstext,amsfonts,amssymb}
\usepackage{epsfig,float}
\usepackage{graphics}
\usepackage{epstopdf}
\usepackage{cite}
\usepackage{psfrag}
\usepackage{subfigure}
\usepackage{url}
\usepackage{shadow,color,pifont,times,rotate}

\newcommand{\mb}{\mathbf}

\newcommand{\mc}{\mathcal}

\def\ex{\times }

\def\esp{{\mathbb E}}  
 \DeclareMathAlphabet{\mathpzc}{OT1}{pzc}{m}{it}
\newtheorem{theorem}{Theorem}[section]

\newtheorem{corollary}{Corollary}
\newtheorem{remark}{Remark}
\newtheorem{definition}{Definition}

\usepackage{enumerate}    
\begin{document}
%

\title{
Broadcasting over the Relay Channel with Oblivious Cooperative Strategy}

\author{
\IEEEauthorblockN{Arash Behboodi and Pablo Piantanida}
\IEEEauthorblockA{Department of Telecommunications, SUPELEC\\
91192 Gif-sur-Yvette, France\\
Email: \{arash.behboodi,pablo.piantanida\}@supelec.fr}
 }

\maketitle
\begin{abstract}
This paper investigates the problem of information transmission over the simultaneous relay channel with two users (or two possible channel outcomes) where for one of them the more suitable strategy is Decode-and-Forward (DF) while for the other one is Compress-and-Forward (CF). In this setting, it is assumed that the source wishes to send common and private informations to each of the users (or channel outcomes). This problem is relevant to: (i) the transmission of information over the broadcast relay channel (BRC) with different relaying strategies and (ii) the transmission of information over the conventional relay channel where the source is oblivious to the coding strategy of relay. A novel coding that integrates simultaneously DF and CF schemes is proposed and an inner bound on the capacity region is derived for the case of general memoryless BRCs.  As special case, the Gaussian BRC is studied where it is shown that by means of the suggested broadcast coding the common rate can be improved compared to existing strategies. Applications of these results arise in broadcast scenarios with relays or in wireless scenarios where the source does not know whether the relay is collocated with the source or with the destination.
\end{abstract}

\thispagestyle{empty}
\section{Introduction}
Cooperation between nodes can serve for boosting the capacity and improving the reliability of the communication, especially in wireless networks. Mainly for this reason, extensive research has been done during the recent years on the topic. The relay channel consists of a sender-receiver pair whose communication is aided by a relay node which helps the communication between the source-destination pair. Substantial advance on this problem was made in \cite{Cover1979}, where upper and inner bounds on the capacity of the discrete memoryless relay channel (DMRC) were established and two cooperative strategies, commonly referred to as \emph{Decode-and-Forward} (DF) and \emph{Compress-and-Forward} (CF), were introduced. Based on these strategies,  further work has been recently done on different aspects of cooperative networks including deterministic channels, and also other examples of cooperative networks like multiple access relay, broadcast relay, multiple relays and fading relay channels, etc. (see \cite{Kramer2005, Liang2007B} and references therein). 

The specification of wireless networks undergoes the extensive changes due to a variety of factors (e.g. interference, fading and user mobility). As a consequence, even when the channels are quasi-static, it is often difficult for the source to know the noise level of the relay link. Hence, the encoder is unable to decide on the suitable coding strategy that would better exploit the presence of the relay. This scenario is frequently seen in \emph{ad-hoc} networks where the source is often assumed to be unaware of the presence of relay users. Nevertheless in most of the previous works the channel is assumed to be fixed and known to all the users. Indeed the problem of uninformed source cooperative networks has been studied in \cite{Katz2005,Behboodi2009}, where achievable rates and coding strategies were developed for relay networks. It is of practical importance to allow the coding to adapt to the channel conditions. In fact, no matter how a set of possible channel outcomes (e.g. level noises, user positions, etc.) can be defined, such scenarios can be addressed as the simultaneous relay channel \cite{Behboodi2009}. In this case, the encoder knows a set of the possible channels but it is unaware of the specific channel that controls the communication. Besides, an interesting connection between simultaneous and broadcast channels (BCs) was first suggested in \cite{coverbroadcast-1972} and then fully exploited for slowly fading MIMO channels \cite{shamai2003}. This idea was used in the context of relay channels in \cite{Maric-Goldsmith2007,Yuksel2004,Behboodi2009,Behboodi2010}.

The performances of DF and CF strategies is directly related to the quality of the channel (e.g. noise conditions) between the relay and the destination. More precisely, DF scheme performs better than CF when the relay is near to the source, whereas CF scheme is more suitable when the relay is near to the destination. In this paper we investigate the simultaneous relay channel (SRC) with two users (or possible channel outcomes). Each of these channels are assumed to be such that in order to take full advantage of the relay, one of them would require to relay the information via DF scheme and the other via CF scheme. This problem can be seen as related to sending common and private informations over the broadcast relay channel (BRC) where one destination is aided by a relay, which uses DF scheme, and the other destination is aided by another relay which uses CF scheme. Based on this approach we derive an inner bound on the capacity region of this scenario. The central idea here is to broadcast information to both users and enabling the source to take simultaneously advantage of DF and CF schemes. It is shown that block Markov coding, commonly used with DF scheme, can be also adapted to CF scheme based on backward decoding idea. Hence when the source sends common information it becomes oblivious to the relaying strategy (similarly to the setting addressed in \cite{Katz2005}, \cite{Katz2006}). 

The organization of this paper is as follows. Section II states definitions along with main results, while the main outlines of the proofs are given in Section III. Section IV provides Gaussian examples and numerical results.

\section{Problem Definitions and Main Results}

\subsection{Problem Definition}
The simultaneous relay channel \cite{Behboodi2009} with discrete source and relay inputs $x\in \mathscr{X}$, $x_T\in \mathscr{X}_T$, discrete channel and relay outputs $y_T\in \mathscr{Y}_T$, $z_T\in \mathscr{Z}_T$, is characterized by two conditional probability distributions (PDs)  $\big\{ P_T:\mathscr{X} \ex \mathscr{X}_T  \longmapsto \mathscr{Y}_T  \ex \mathscr{Z}_T\big\}_{T= 1,2}$, where $T$ is the channel index. It is assumed here that the transmitter (the source) is unaware of the realization of  $T$ that governs the communication, but $T$ should not change during the communication. However, $T$ is assumed to be known at the destination and the relay ends. 

\begin{definition}[Code] \label{def-code}
A code for the SRC consists of: (i) an encoder mapping $\{ \varphi:\mc{W}_{1} \ex\mc{W}_{2}  \longmapsto \mathscr{X}^n  \}$, (ii) two decoder mappings $\{ \psi_T:\mathscr{Y}_T^n \longmapsto \mc{W}_{T} \}$ and (iii) a set of relay functions $\{f_{T,i}\}_{i=1}^n$ such that $\{ f_{T,i} :\mathscr{Z}_T^{i-1}  \longmapsto \mathscr{X}_T^n  \}_{i=1}^n$, for some finite sets of integers $\mc{W}_{T}=\big\{ 1,\dots, W_{T} \big\}$. The rates of such code are $n^{-1} \log W_{T}$ and its maximum error probability 
\begin{equation*}
e_{\max,T}^{(n)} \doteq \max_{(w_0,w_T)\in\mc{W}_{0n}\times\mc{W}_{T}} \Pr \Big\{  \psi (\mb{Y} _T ) \neq (w_0,w_T)  \Big\}.\label{errorprob_def}    
\end{equation*}    
\end{definition}

\begin{definition}[Achievable rates and capacity] 
For every $0$ $ < \epsilon, \gamma< 1$, a triple of non-negative numbers $(R_0,R_1,R_2)$ is achievable for the SRC if for every sufficiently large $n$ there exist $n$-length block code whose error probability satisfies $\max_{T=\{1,2\}} e_{\max,T}^{(n)}\big( \varphi,\psi, \{f_{T,i}\}_{i=1}^n\big) \leq \epsilon$  and the rates 
$n^{-1} \log W_{T} \geq R_T-\gamma$ for each $T=\{0,1,2\}$. The set of all achievable rates is called the capacity region for the SRC. We emphasize that no prior distribution on $T$ is assumed and thus the encoder must exhibit a code that yields small error probability for every $T=\{1,2\}$, yielding the BRC setting. A similar definition can be offered for the common-message BC with a single message set $\mc{W}_{0}$ and rate $n^{-1} \log W_{0}$. 
\end{definition}

\begin{figure}[t]
\centering
\subfigure[State-Dependent Relay Channel $T$]{
	\includegraphics [scale=0.5] {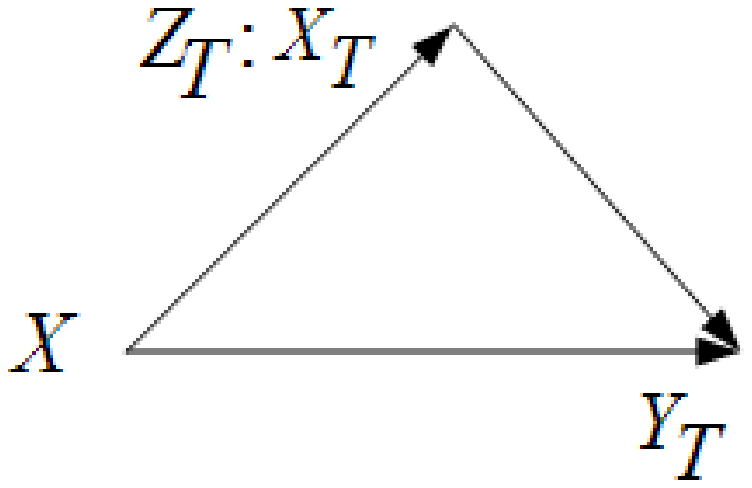}
\label{fig:1}
}
\subfigure[Broadcast Relay Channel]{
	\includegraphics [scale=0.4] {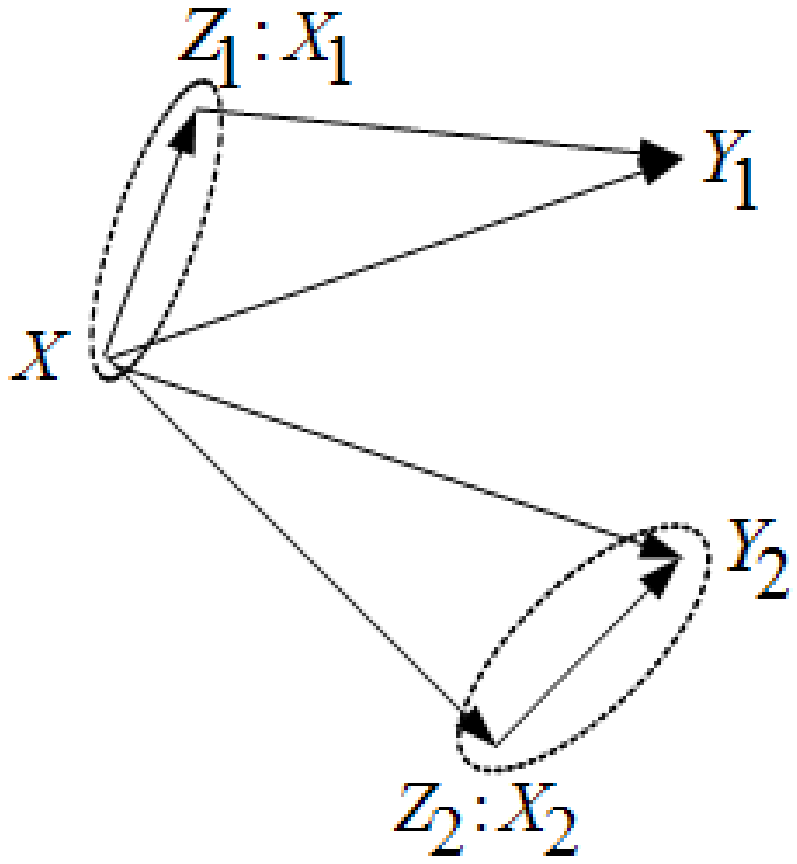}
\label{fig:2}
}
	\caption{Simultaneous Relay Channel.}       
 
\end{figure}

Since the relay and the receiver can be assumed to be cognizant of the realization $T$, the problem of coding for the SRC can be turned into that of the BRC \cite{Behboodi2009}. This consists of two relay branches where each one equals to a relay channel with $T=\{1,2\}$, as is shown in Fig. \ref{fig:2}.  The encoder sends common and private messages  $(W_0,W_T)$ to destination $T$ at rates $(R_0,R_T)$. The BRC is defined by the PD $\big\{P:\mathscr{X} \ex \mathscr{X}_1  \ex \mathscr{X}_2  \longmapsto \mathscr{Y}_1  \ex \mathscr{Z}_1 \ex \mathscr{Y}_2  \ex \mathscr{Z}_2 \big\}$, with channel and relay inputs $(x,x_1,x_2)$ and channel and relay outputs  $(y_1,z_1,y_2,z_2)$. Notions of achievability for $(R_0,R_1,R_2)$ and capacity remain the same as for BCs (see \cite{coverbroadcast-1972}, \cite{Kramer2005} and \cite{Liang2007A}). 

\subsection{Coding Theorem for the Broadcast Relay Channel}
\begin{theorem} An inner bound on the capacity region of the BRC with oblivious cooperative strategy is given by
\begin{align*}
\mathscr{R}_{DF-CF}  \doteq  \displaystyle{\bigcup\limits_{P\in\mathscr{P}}}  \Big\{(&R_0\geq 0, R_1\geq 0,R_2 \geq 0):  \\ 
R_0+R_1 & \leq I_1  \\
R_0+R_2 & \leq I_2-I(U_2;X_1\vert U_0V_0)  \\
R_0+R_1+R_2 &\leq I_1+J_2-I(U_1X_1;U_2\vert U_0V_0) \\
R_0+R_1+R_2 &\leq J_1+I_2-I(U_1X_1;U_2\vert U_0V_0) \\   
2R_0+R_1+R_2 & \leq I_1+I_2-I(U_1X_1;U_2\vert U_0V_0) \Big\},\label{eq:II-1}
\end{align*}
where the quantities $(I_i,J_i)$ with $i=\{1,2\}$ are given by 
\begin{equation*}
\begin{array} {l}
I_1\doteq \min\big\{I(U_0U_1;Z_1\vert X_1V_0),I(U_1U_0X_1V_0;Y_1)\big\},\\
I_2\doteq I(U_2U_0V_0;\hat{Z}_2Y_2\vert X_2),   \\
J_1\doteq \min\big\{I(U_1;Z_1\vert X_1U_0V_0),I(U_1X_1;Y_1\vert U_0V_0)\big\},\\
J_2\doteq I(U_2;\hat{Z}_2Y_2\vert X_2U_0V_0),   \\
\end{array}   
\end{equation*}
and  the set of all admissible PDs $\mathscr{P}$ is defined as 
\begin{equation*}
\begin{array}{l}
\mathscr{P}\doteq \big \{P_{V_0U_0U_1U_2X_1X_2XY_1Y_2Z_1Z_2\hat{Z}_2}= P_{V_0}P_{X_2}P_{X_1|V_0}\\P_{U_0|V_0} P_{U_2U_1|X_1U_0} P_{X|U_2U_1}P_{Y_1Y_2Z_1Z_2|XX_1X_2}P_{\hat{Z}_2|X_2Z_2},   \\
 I(X_2;Y_2) \geq I(Z_2;\hat{Z}_2\vert X_2Y_2), \,\, \\
 (V_0,U_0,U_1,U_2) \minuso (X_1,X_2,X) \minuso (Y_1,Z_1,Y_2,Z_2) \big \}. \nonumber
\end{array}
\label{eq:II-2}
\end{equation*}  
\label{thm:1}
\end{theorem}
\begin{corollary}[common-information]
An lower bound on the capacity of the common-message BRC is given by   
\begin{equation*}
\begin{array} {l}
R_0 \leq \max\limits_{P_{X_1X_2X}\in\mathscr{P}}\min\big\{I(X;Z_1\vert X_1),I(X,X_1;Y_1),\\
\,\,\,\,\,\,\,\,\,\,\,\,\,\,\,\,\,\,\,\,\,\,\,\,\,\,\,\,\,\,\,\,\,\,\,\,\,\,\,\,\,\,\,\,\,\,\,\,\,\,\,\,\,\,\,\,\,\,\, \,\,\,\,\,\,\,\,\,\,\,I(X;\hat{Z}_2Y_2\vert X_2)\big\}.
\end{array}    
\label{eq:II-3}
\end{equation*}
\label{thm:2}
\end{corollary}

\begin{corollary}[private information]   An inner bound on the capacity region of the BRC with heterogeneous cooperative strategies is given by the set of rates $(R_1,R_2)$ satisfying 
\begin{align*}
R_1 & \leq \min\big\{I(U_1;Z_1\vert X_1),I(U_1X_1;Y_1)\big\} \\
R_2 & \leq I(U_2;\hat{Z}_2Y_2\vert X_2)-I(U_2;X_1) \\
R_1+R_2 & \leq \min\big\{I(U_1;Z_1\vert X_1),I(U_1X_1;Y_1)\big\} \\
& + I(U_2;\hat{Z}_2Y_2\vert X_2) -I(U_1X_1;U_2),
\label{eq:II-4}
\end{align*}
for all joint PDs $P_{U_1U_2 X_1X_2XY_1Y_2Z_1Z_2\hat{Z}_2} \in \mathscr{P}$. 
\label{thm:3}
\end{corollary}

\begin{remark}
The region in theorem \ref{thm:1} includes Marton's region \cite{Marton1979} with $(X_1,X_2,V_0)=\emptyset$, $Z_1=Y_1$ and $Z_2=Y_2$.  

Observe that the rate corresponding to the DF scheme that appears in theorem \ref{thm:2} coincides with the conventional DF rate. Here a common code for DF and CF users is employed hence it shows that a block Markov coding is essentially oblivious to the relaying strategy and proves the same performance in both cases. An outline of the proof is given in Section III.
Corollary \ref{thm:2} follows by choosing $U_1=U_2=\emptyset$, $V_0=X_1$ and $U_0=X$. Whereas corollary \ref{thm:3} follows by setting $U_0=V_0=\emptyset$. The next theorem presents an upper bound on capacity of the common-message BRC.
 \end{remark}

\begin{theorem}[upper bound]
An upper bound on the capacity region of the common-message BRC is given by
\begin{equation*}
\begin{array} {l}
R_0 \leq \max\limits_{P_{X_1X_2X}\in\mathscr{P} }\min\big\{I(X;Z_1Y_1\vert X_1),I(X,X_1;Y_1),\\
\,\,\,\,\,\,\,\,\,\,\,\,\,\,\,\,\,\,\,\,\,\,\,\,\,\,\,\,\,\,\,\,\,\,\,\,\,\,\,\,\,\,\,\,\,\,\,\,\,\,\,\,\,\,\,\,\,\,\, \,\,\,\,\,\,\,\,\,\,\,I(X;Z_2Y_2\vert X_2),I(X,X_2;Y_2)\big\}.
\end{array}
\label{eq:II-5}
\end{equation*}
\label{thm:4}
\end{theorem}

\section{Sketch of the Proofs}
\subsection{Proof of Theorem \ref{thm:1}}

\begin{figure} [t]
\centering   
\includegraphics [width=0.45\textwidth] {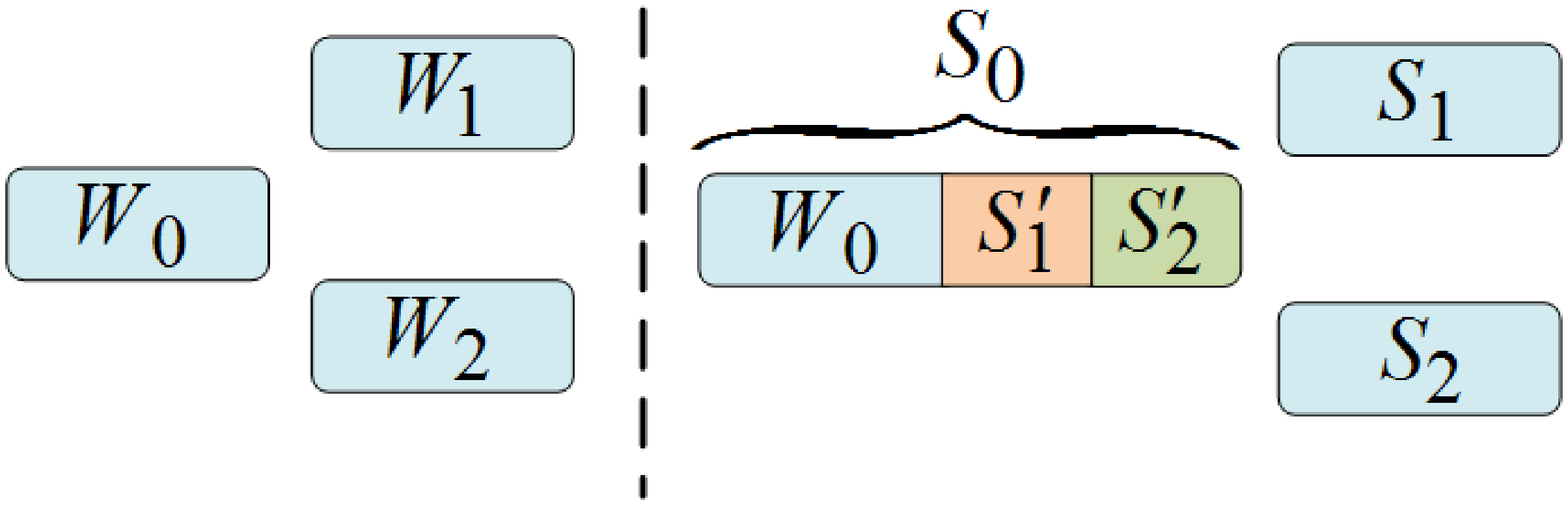}   
\caption{Message reorganization}
\label{fig:reorg}
\end{figure}

Reorganize first private messages, Fig \ref{fig:reorg} $w_i$, $i\in \{1,2\}$ into $({s'}_i,{s}_i)$ with non-negative rates $(S'_i,S_i)$ where $R_i=S'_i+S_i$. Merge $({s'}_1,{s'}_2,w_0)$ to one message ${s}_0$ with rate $S_0=R_0+S'_1+S'_2$. 
\textit{Code Generation:}
\begin{enumerate} [(i)]
	\item Randomly and independently generate $2^{nS_0}$ sequences $\underline{v}_0$ draw i.i.d. from $P_{V_0}(\underline{v}_0)=\prod_{j=1}^np_{V_0}(v_{0j})$. 
	Index them as $\underline{v}_0(r_0)$ with $r_0\in \left[1,2^{nS_0}\right]$.
	\item 
	For each $\underline{v}_0(r_0)$, randomly and independently generate $2^{nS_0}$ sequences $\underline{u}_0$ draw i.i.d. from $P_{U_0|V_0}(\underline{u}_0\vert \underline{v}_0(r_0))=\prod_{j=1}^np_{U_0\vert V_0}(u_{0j}\vert v_{0j}(r_0))$. 
        Index them as $\underline{u}_0(r_0,s_0)$ with $s_0\in \left[1,2^{nS_0}\right]$.
	\item
	For each $\underline{v}_0(r_0)$, randomly and independently generate $2^{nT_1}$ sequences $\underline{x}_1$ draw i.i.d. from $P_{X_1|V_0}(\underline{x}_1\vert \underline{v}_0(r_0))=\prod_{j=1}^n p_{X_1\vert V_0}(x_{1j}\vert v_{0j}(r_0))$.
	Index them as $\underline{x}_1(r_0,r_1)$ with $r_1\in \left[1,2^{nT_1}\right]$.
\item 
	Randomly and independently generate  $2^{n{R}_{x_2}}$ sequences $\underline{x}_2$ draw i.i.d. from $P_{X_2}(\underline{x}_2)=\prod_{j=1}^np_{X_2}(x_{2j})$ as $\underline{x}_2(r_2)$, where $r_2\in \left[1,2^{n{R}_{x_2}}\right]$.
	\item
	For each $\underline{x}_2(r_2)$ randomly and independently generate $2^{n\hat{R}_2}$ sequences  $\underline{\hat{z}}_2$ each with probability $P_{\hat{Z}_2|X_2}(\underline{\hat{z}}_2\vert \underline{x}_2(r_2))= \prod_{j=1}^np_{\hat{Z}_2\vert X_2}(\hat{z}_{2j}\vert x_{2j}(r_2))$. Index them as $\underline{\hat{z}}_2(r_2,\hat{s})$, where $\hat{s}\in \left[1,2^{n\hat{R}_2}\right]$.
\item   
	Partition the set $\big\{1,\ldots,2^{n\hat{R}_2}\big\}$ into $2^{nR_2}$ cells and label them as $S_{r_2}$. In each cell there are $2^{n(\hat{R}_2-R_2)}$ elements. 
	\item
	For each pair $\big(\underline{u}_0(r_{0},s_{0}),\underline{x}_1(r_{0},r_{1})\big)$, randomly and independently generate $2^{nT_1}$ sequences $\underline{u}_{1}$ draw i.i.d. from   $P_{U_1|U_0X_1V_0}(\underline{u}_{1}\vert \underline{u}_0(r_{0},s_{0}),\underline{x}_1(r_{0},r_{1}),\underline{v}_0(r_{0}))=$\\
	$\prod_{j=1}^np_{U_1\vert U_0V_0X_1}(u_{1j}\vert u_{0j}(r_{0},s_{0}),x_{1j}(r_{0},r_{1}),v_{0j}(r_{0})) $. Index them as $\underline{u}_1(r_{0},s_{0},r_{1},t_{1})$, where $t_1\in \left[1,2^{nT_1}\right]$.
\item 
	For each $\underline{u}_0(r_{0},s_{0})$, randomly and independently generate $2^{nT_2}$ sequences $\underline{u}_{2}$  draw i.i.d. from  $P_{U_2| U_0V_0}(\underline{u}_{2}\vert  \underline{u}_0(r_{0},s_{0}),\underline{v}_0(r_{0}))=$ \\ $ \prod_{j=1}^np_{U_2\vert U_0V_0}(u_{2j}\vert u_{0j}(r_{0},s_{0}),v_{0j}(r_{0})) $. Index them as $\underline{u}_2(r_{0},s_{0},t_{2})$, where $t_2\in \left[1,2^{nT_2}\right]$.
	\item
For $b\in\{1,2\}$, partition the set $\big\{1,\ldots,2^{nT_b}\big\}$ into $2^{nS_b}$ subsets and label them as $S_{s_b}$. In each subset, there are $2^{n(T_b-S_b)}$ elements.
\item \label{FirstMarton}
Then for each subset $S_{s_{2}}$, create the set $\mathscr{L}$ consisting of those index $t_2$ such that $t_2\in S_{s_{2}}$, and  $\underline{u}_2\big(r_{0},s_{0},t_{2}\big)$ is jointly typical with $\underline{x}_{1}\big(r_{0}, r_{1}\big),\underline{v}_0\big(r_{0}\big), \underline{u}_0\big(r_{0},s_{0}\big)$. 
\item  \label{SecondMarton}
Then it looks for $t_1\in S_{s_{1}}$ and $t_2\in\mathscr{L}$ such that $\big(\underline{u}_1(r_{0},s_{0},r_{1},t_{1})$,$ \underline{u}_2(r_{0},s_{0},t_{2})\big)$ are jointly typical given the RVs $\underline{v}_0(r_{0}), \underline{x}_1(r_{0},r_{1}), $ and with $\underline{u}_0(r_{0},s_{0})$. The constraints for the coding steps (\ref{FirstMarton}),(\ref{SecondMarton}) are: 
\begin{align}
T_2-S_2 & \geq I(U_2;X_1\vert U_0V_0), \label{eq:III-1A}\\
T_1+T_2-S_1-S_2 & \geq I(U_2;U_1X_1\vert U_0V_0). \label{eq:III-1B}
\end{align}
The first inequality guarantees the existence of non-empty sets $\mathscr{L}$. 
\item
Finally, use a deterministic function for generating $\underline{x}$ as $f\left(\underline{u}_1,\underline{u}_2\right)$ indexed by $\underline{x}(r_{0},s_{0},r_{1},t_{1},t_{2})$.
\end{enumerate}
\textit{Encoding Part:} In block $i$, the source wants to send $(w_{0i},w_{1i},w_{2i})$ by reorganizing them into $(s_{0i},s_{1i},s_{2i})$. Encoding steps are as follows:
\begin{enumerate}[(i)]
\item
Relay $1$ knows supposedly $(s_{0(i-1)},t_{1(i-1)})$ so it sends $\underline{x}_1\big(s_{0(i-1)},t_{1(i-1)}\big)$.
\item
Relay $2$ knows from the previous block that $\hat{s}_{i-1} \in S_{r_{2i}}$ and it sends $\underline{x}_2(r_{2i})$.
\item
From $(s_{0i},s_{1i},s_{2i})$, the source finds $(t_{1i},t_{2i})$ and sends $\underline{x}(s_{0(i-1)},s_{0i},t_{1(i-1)},t_{1i},t_{2i})$. 
\end{enumerate}
\textit{Decoding Part:} After the transmission of the block $i+1$, the first relay starts to decode the messages of block $i+1$ with the assumption that all messages up to block $i$ have been correctly decoded. Destination 1 waits until the last block and uses backward decoding (similarly to \cite{Kramer2005}). The second destination first decodes $\hat{Z}_2$ and then uses it with $Y_2$ to decode the messages, while the second relay tries to find $\hat{Z}_2$. 
\begin{enumerate}[(i)]
\item
Relay $1$ tries to decode $(s_{0(i+1)},t_{1(i+1)})$ subject to:
\begin{align}
T_1+S_0 & < I(U_0U_1;Z_1\vert X_1V_0), \label{eq:III-2A}    \\
T_1 & < I(U_1;Z_1\vert U_0V_0X_1). 
\label{eq:III-2B}
\end{align}
\item
Destination $1$ tries to decode $(s_{0i},t_{1i})$ subject to
\begin{align}
T_1+S_0 & < I(X_1V_0U_0U_1;Y_1),  \label{eq:III-3A} \\
T_1 & < I(U_1X_1;Y_1\vert U_0V_0). 
\label{eq:III-3B}
\end{align}
\item 
Relay $2$ searches for $\hat{s}_{i}$ after receiving $\underline{z}_2(i)$ such that $\big(\underline{x}_{2}(r_{2i}),\underline{z}_{2}(i),\underline{\hat{z}}_{2}(\hat{s}_{i},r_{2i})\big)$ are jointly typical subject to
\begin{equation}
\hat{R}_2 \geq I(Z_2;\hat{Z}_2|X_2).
\label{eq:III-4}
\end{equation}
\item 
Destination $2$ searches for $r_{2(i+1)}$ such that $\big(\underline{y}_2(i+1),\underline{x}_{2}(r_{2(i+1)})\big)$ are jointly typical. Then it finds  $\hat{s}_{i}$ such that $\hat{s}_{i}\in S_{r_{2(i+1)}}$ and $\big(\underline{\hat{z}}_{2}(\hat{s}_{i},r_{2i}),\underline{y}_2(i),\underline{x}_{2}(r_{2i})\big)$ are jointly typical. Conditions for reliable decoding are
\begin{equation}
\begin{array} {l}
R_{x_2} \leq I(X_2;Y_2),\,\,\hat{R}_2 \leq R_{x_2}+I(\hat{Z}_2;Y_2|X_2).
\end{array}
\label{eq:III-5}
\end{equation}
\item
Decoding of CF user in block $i$ is done with the assumption of correct decoding of $(s_{0l},t_{2l})$ for $l\leq i-1$. The  pair $(s_{0i},t_{2i})$ are decoded as the message such that $(\underline{v_0}(s_{0(i-1)}),\underline{u_0}(s_{0(i-1)},s_{0i}),\underline{u_2}(s_{0(i-1)},s_{0i},t_{2i}),\underline{y}_{2}(i)$\\$,\underline{\hat{z}}_2(\hat{s}_i,,r_{2i}),\underline{x}_{2}(r_{2i}))$ and $(\underline{v}_0(s_{0i}),\underline{y}_{2}(i+1),$\\$\underline{\hat{z}}_2(\hat{s}_{i+1},r_{2(i+1)}),\underline{x}_{2}(r_{2(i+1)}))$ are all  jointly typical. This leads to the next constraints
\begin{align}
S_0+T_2 & \leq I(V_0U_0U_2;Y_2\hat{Z}_2\vert X_2),\label{eq:III-6A}\\
T_2 & \leq I(U_2;Y_2\hat{Z}_2\vert V_0U_0X_2). 
\label{eq:III-6B}
\end{align}
It is interesting to see that regular coding allows us to use the same code for DF and CF scenarios, while keeping the same final CF rate.                                      
\end{enumerate}
After decoding $(s_{0i},s_{1i},s_{2i})$ at destinations, the original messages $(w_{0i},w_{1i},w_{2i})$ can be extracted. One can see that the rate region of theorem \ref{thm:1} follows form equations \eqref{eq:III-1A}-\eqref{eq:III-6B}, the equalities between the original rates and reorganized rates, the fact that all the rates are positive and by finally using Fourier-Motzkin elimination. Similar to \cite{Cover1979}, the necessary condition $I(X_2;Y_2)\geq I(Z_2;\hat{Z}_2\vert X_2Y_2)$ follows from \eqref{eq:III-4} and \eqref{eq:III-5}.
\subsection{Proof of Theorem \ref{thm:4}}
It is easy to show that the upper bounds presented in theorem \ref{thm:4} are a combination of two cut-set bounds for the relay channel. 

\section{Application Example: The Gaussian BRC}
In this section the Gaussian BRC is analyzed, where the relay is collocated with the source in the first channel and with the destination in the second one, as is shown in Fig. \ref{fig:2}. No interference is allowed from the relay $b$ to the destination $\overline{b}$, $b\in\{1,2\}$. The relationship between RVs are as follows: 
\begin{equation*}
\begin{array} {c }
Y_1=\frac{X}{\sqrt{d^\delta_{y_1}}}+\frac{X_1}{\sqrt{d^\delta_{z_1y_1}}}+\mathpzc{N}_1\,\,,\,\, Y_2=\frac{X}{\sqrt{d^\delta_{y_2}}}+\frac{X_2}{\sqrt{d^\delta_{z_2y_2}}}+\mathpzc{N}_2,\\
Z_1=\frac{X}{\sqrt{d^\delta_{z_1}}}+\tilde{\mathpzc{N}}_1\,\,,\,\, Z_2=\frac{X}{\sqrt{d^\delta_{z_2}}}+\tilde{\mathpzc{N}}_2.
\end{array}
\label{eq:9}
\end{equation*}
The channel inputs are constrained to satisfy the power constraint $P$, while the relay inputs must satisfy power constraint $P_1$ and $P_2$. The Gaussian noises $\widehat{\mathpzc{N}}_2,\tilde{\mathpzc{N}}_1,\tilde{\mathpzc{N}}_2$, ${\mathpzc{N}}_1$ and ${\mathpzc{N}}_2$  are zero-mean of variances  $\tilde{{N}}_1,\tilde{{N}}_2,\widehat{N}_2,N_1$ and $N_2$. 

\subsection{Achievable rates for private information}
Based on Corollary \ref{thm:3} we derive achievable rates for the case of private information. As for the classical broadcast channel, by using superposition coding, we decompose $X$ as a sum of two independent RVs such that $\esp\left\{X_A^2\right\}=\alpha P$ and $\esp\left\{X_B^2\right\}=\overline{\alpha} P$, where $\overline{\alpha}=1-\alpha$.  The codewords $(X_A,X_B)$ contain the information intended to receivers $Y_1$ and $Y_2$. We choose also $\hat{Z}_2=Z_2+\hat{\mathpzc{N}}_2$. First, we identify two different cases for which dirty-paper coding (DPC) schemes are derived.  

\textit{Case I:}
A DPC scheme is applied to $X_B$ for canceling the interference  $X_A$, while for the relay branch of the channel is similar to \cite{Cover1979}. Hence, the auxiliary RVs $(U_1,U_2)$ are set to
\begin{equation}
U_1=X_A=\tilde{X}_A+\sqrt{\frac{\overline{\beta}\alpha P}{P_1}}X_1,\,\,\,\, U_2=X_B+\gamma X_A,
\label{eq:10}
\end{equation}
where $\beta$ is the correlation coefficient between the relay and source, and $\tilde{X}_A$ and $X_1$ are independent. Notice that  in this case, instead of only $Y_2$, we have also $\hat{Z}_2$ present in the rate. Thus DPC should be also able to cancel the interference in both, received and compressed signals which have different noise levels. Calculation should be done again with $(Y_2,\hat{Z}_2)$ which are the main message $X_B$ and the interference $X_A$. We can show that the optimum $\gamma$ has a similar form to the classical DPC with the noise term replaced by an equivalent noise which is like the harmonic mean of the noise in $(Y_2,\hat{Z}_2)$. The optimum $\gamma^*$ is given by $\frac{\overline{\alpha}P}{\overline{\alpha}P+N_{t1}}$ where $N_{t1}=\big( (d^\delta_{z_2}(\tilde{N}_2+\widehat{N}_2))^{-1}+(d^\delta_{y_2}({N}_2))^{-1}\big)^{-1}$. As we can see the equivalent noise is twice of the harmonic mean of the other noise terms. 

From corollary \ref{thm:3} we can see that the current definitions yield the rates: $R_1=\min\big\{I(U_1;Z_1\vert X_1),I(U_1X_1;Y_1)\big\}$ and $R_2= I(U_2;Y_2\hat{Z}_2\vert X_2)-I(U_1X_1;U_2)$. The rate for optimal $\gamma$ is as follows: 
\begin{align}
R_1^*=&\underset{0\leq\beta\leq1}{\max} \min\Big\{C\left(\frac{\alpha \frac{P}{d^\delta_{y_1}}+\frac{P_1}{d^\delta_{z_1y_1}}+2\sqrt{\frac{\overline\beta\alpha PP_1}{d^\delta_{y_1}d^\delta_{z_1y_1}}}}{\frac{\overline\alpha P}{d^\delta_{y_1}}+{N}_1}\right), \nonumber \\
&\,\,\,\,\,\,\,\,\,\,\,\,\,\,\,\,\,\,\,\,\,\,\,\,\,\,\,\,\,\,\,\,
C\left(\frac{\alpha\beta P}{\overline\alpha P+d^\delta_{z_1}\tilde{N}_1}\right)\Big\},\nonumber \\
R_2^*=&C\left(\frac{\overline{\alpha}P}{d^\delta_{y_2}N_2}+\frac{\overline{\alpha}P}{d^\delta_{z_2}(\widehat{N}_2+\tilde{N}_2)}\right),
\label{eq:11}
\end{align}
where $C(x)=\frac{1}{2}\log(1+x)$. Note that since $(X_A,X_B)$ are chosen independent, destination 1 sees $X_B$ as an additional channel noise. The compression noise is chosen as follows
\begin{equation}
\widehat{N}_2=\tilde{N}_2{\left(P\left(\frac{1}{d^\delta_{y_2}N_2}+\frac{1}{d^\delta_{z_2}\tilde{N}_2}\right)+1\right)}/{\frac{P_2}{d^\delta_{z_2y_2}N_2}}.
\label{eq:12}
\end{equation}

\textit{Case 2:} We use a DPC scheme for $Y_2$ to cancel the interference $X_1$, and next we use a DPC scheme for $Y_1$ to cancel $X_B$. For this case, the auxiliary RVs $(U_1,U_2)$ are
\begin{equation}
\left\{\begin{array} {l}
U_1=X_A+\lambda\,\, X_B \textrm{ with }\,\, X_A=\tilde{X}_A+\sqrt{\frac{\overline{\beta}\alpha P}{P_1}}X_1, \\
U_2=X_B+\gamma X_1.
\end{array} \right.
\label{eq:13}
\end{equation}
From corollary \ref{thm:3} the rates with the current definitions are $R_1=\min\big\{I(U_1;Z_1\vert X_1),I(U_1X_1;Y_1)\big\}-I(U_1;U_2|X_1)$ and $R_2= I(U_2;Y_2\hat{Z}_2\vert X_2)-I(X_1;U_2)$. The argument for the destination 2  is similar but it differs in its DPC, since only $X_1$ can be canceled and $X_A$ appears as additional noise. The optimum $\gamma^*$ similar to \cite{Behboodi2009} will be $\gamma^*=\sqrt{\frac{\overline{\beta}\alpha P}{P_1}}\frac{\overline{\alpha}P}{\overline{\alpha}P+N_{t2}}$
where $N_{t2}=\big( (d^\delta_{z_2}(\tilde{N}_2+\widehat{N}_2)+{\beta}\alpha P)^{-1}+(d^\delta_{y_2}({N}_2)+{\beta}\alpha P)^{-1}\big)^{-1}$,
\begin{equation}
R_2^*=C\left(\frac{\overline{\alpha}P}{d^\delta_{y_2}N_2+{\beta}\alpha P}+\frac{\overline{\alpha}P}{d^\delta_{z_2}(\widehat{N}_2+\tilde{N}_2)+{\beta}\alpha P}\right).
\label{eq:15}
\end{equation}

For destination 1, the achievable rate is the minimum of two mutual informations, where the first term is given by $R_{11}= I(U_1;Z_1\vert X_1)-I(U_1;U_2\vert X_1)$ that yields 
\begin{equation}
\begin{array} {l}
R_{11}^{(\beta,\lambda)}=\frac{1}{2}\log\left(\frac{\alpha\beta P(\alpha\beta P+\overline{\alpha}P+d^\delta_{z_1}\tilde{N}_1)}
{d^\delta_{z_1}\tilde{N}_1(\alpha\beta P+\lambda^2\overline{\alpha}P)+(1-\lambda)^2\overline{\alpha}P\alpha{\beta}P}\right).\\
\end{array}
\label{eq:16}
\end{equation}
The second term is $R_{12}=I(U_1X_1;Y_1)-I(U_1;U_2\vert X_1)$, where the first mutual information can be decomposed into two terms $I(X_1;Y_1)$ and $I(U_1;Y_1\vert X_1)$. 
Notice that regardless of the former, the rest of the terms in the expression of rate $R_{12}$ are similar to $R_{11}$. 
The main codeword is $\tilde{X}_A$, while $X_B$ and ${\mathpzc{N}}_1$ represent  the random state and the noise, respectively. After adding the term $I(X_1;Y_1)$ we have
\begin{equation}
\begin{array} {l}
R_{12}^{(\beta,\lambda)}=\frac{1}{2}\log\left(\frac{\alpha\beta P(\frac{P}{d^\delta_{y_1}}+\frac{P_1}{d^\delta_{z_1y_1}}+2\sqrt{\frac{\overline{\beta}\alpha PP_1}{d^\delta_{y_1}d^\delta_{z_1y_1}}}+N_1)}
{{N}_1(\alpha\beta P+\lambda^2\overline{\alpha}P)+(1-\lambda)^2\frac{\overline{\alpha}P\alpha{\beta}P}{d^\delta_{y_1}}}\right).\\
\end{array}
\label{eq:17}
\end{equation}
Based on expressions \eqref{eq:17} and \eqref{eq:16}, the maximum achievable rate follows as 
\begin{equation}
R^*_1=\underset{0\leq\beta,\lambda\leq1}{\max} \min\big\{R_{11}^{(\beta,\lambda)},R_{12}^{(\beta,\lambda)}\big\}.   \label{eq:17B}
\end{equation}
It should be noted that the constraints for $\widehat{N}_2$ is still the same as \eqref{eq:12}.

 \begin{figure} [t]
\centering  
\includegraphics [width=0.50\textwidth] {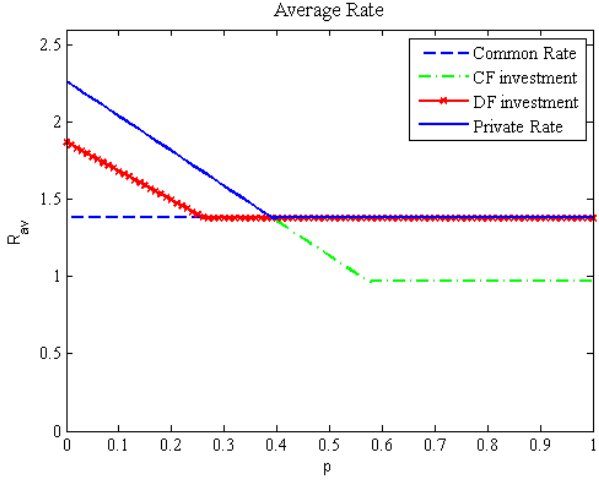}    
\caption{Expected rate of the Gaussian random relay channel}
\label{fig:3}
\end{figure}

\subsection{Lower and upper bounds on the common-rate}
Based on Corollary \ref{thm:2} and Theorem \ref{thm:4} we derive lower and upper bounds on the common-rate. The definition of the channels remain the same. We define $X=U+\sqrt{\frac{\overline{\beta}P}{P_1}}X_1$ and evaluate corollary \ref{thm:2}. The goal is to send common-information at rate $R_0$. It is easy to verify that the two DF rates in corollary \ref{thm:2} result in the classical rates \cite{Kramer2005}:
\begin{align}
R_0\leq& C\left(\frac{\beta P}{d^\delta_{z_1}\tilde{N}_1}\right), \label{eq:18A}\\
R_0\leq& C\left(\frac{\frac{P}{d^\delta_{y_1}}+\frac{P_1}{d^\delta_{z_1y_1}}+2\sqrt{\frac{\overline\beta PP_1}{d^\delta_{y_1}d^\delta_{z_1y_1}}}}{{N}_1}\right). 
\label{eq:18B}
\end{align}  \vspace{-1mm}
Whereas the CF rate given by $I(UX_1;Y_2\hat{Z}_2|X_2)$ is as follows
\begin{equation}
R_0\leq C\left(\frac{P}{d^\delta_{y_2}N_2}+\frac{P}{d^\delta_{z_2}(\widehat{N}_2+\tilde{N}_2)}\right).
\label{eq:19}
\end{equation}
And the upper bound from theorem \ref{thm:4} is given as \vspace{-1mm}
\begin{equation*}
\!\!\!\!\!\!  \begin{array}{l}
R^*_1=\underset{0\leq\beta_1,\beta_2\leq1}{\max} \min \\
\Big\{ C\left(\beta_1 P\left[\frac{1}{d^\delta_{z_1}\tilde{N}_1}+\frac{1}{d^\delta_{y_1}{N}_1}\right]\right),
C\left(\frac{\frac{P}{d^\delta_{y_1}}+\frac{P_1}{d^\delta_{z_1y_1}}+2\sqrt{\frac{\overline\beta_1 PP_1}{d^\delta_{y_1}d^\delta_{z_1y_1}}}}{{N}_1}\right),\\
C\left(\beta_2 P\left[\frac{1}{d^\delta_{z_2}\tilde{N}_2}+\frac{1}{d^\delta_{y_2}{N}_2}\right]\right),
C\left(\frac{\frac{P}{d^\delta_{y_2}}+\frac{P_2}{d^\delta_{z_2y_2}}+2\sqrt{\frac{\overline\beta_2 PP_2}{d^\delta_{y_2}d^\delta_{z_2y_2}}}}{{N}_2}\right)
\Big\}. 
\end{array}
\end{equation*}
\begin{remark}
Observe that the rate \eqref{eq:19} is exactly the same as classical Gaussian CF \cite{Kramer2005}. This means that DF regular encoding can also be decoded in CF channel, as well for the case with collocated relay and receiver. The constraint for the compression noise remains unchanged.  Notice that this observation parallels that made in recent results  reported in \cite{Katz2005}  for a single relay channel, where the source is obviously to the presence of the relay.  In our setting, the source is obviously to the relaying scheme of the relay. 
\end{remark} 
\subsection{Numerical Results and Discussion}
\begin{figure} [t]
\centering   
\includegraphics [width=0.49\textwidth] {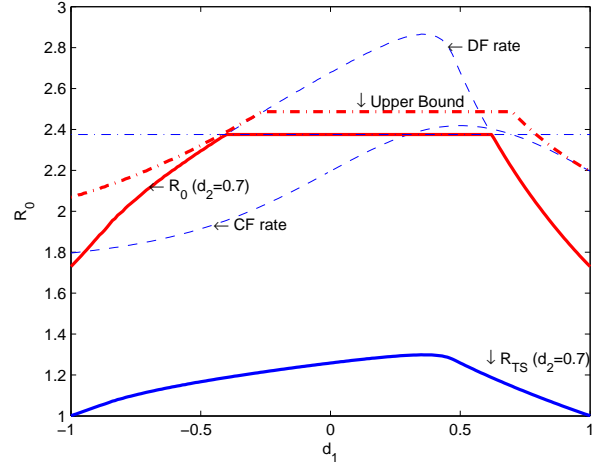}   
\caption{Common rate of the Gaussian BRC}
\label{fig:4}
\end{figure}

In the previous sections we showed that by using the proposed coding it is possible to send common information at the minimum rate between CF and DF schemes $R_0=\min\{R_{DF},R_{CF}\}$ (i.e. expressions \eqref{eq:18A} to \eqref{eq:19}).  For the case of private information, we showed that any pair of rates $(R_{DF}\leq R^*_1,R_{CF}\leq R^*_2)$ given by  \eqref{eq:15} and \eqref{eq:17B} are admissible and thus $(R_{DF},R_{CF})$ can be simultaneously sent. 

We now assume a composite model where the relay is collocated with the source with probability $p$ (refer to it as the first channel) and with the destination with probability $1-p$ (refer to it as the second channel). Therefore DF scheme is the suitable strategy for the first channel while CF scheme performs better on the second one. Then for any triple of achievable rates $(R_0,R_1,R_2)$ we define the expected rate as 
$$
R_{av}=R_0+pR_{1}+(1-p)R_{2}.
$$ 
The expected rates achieved with the proposed coding strategy and via conventional strategies  are compared. Alternative coding schemes for this scenario are possible. The encoder can simply invest on one coding scheme DF or CF, which is useful when the probability of one channel is high so the source invests only on it. In fact, there are different ways to proceed: (i) Send information via DF scheme at the best possible rate between both channels. Then the worst channel cannot decode and thus the expected rate is $p^{\max}_{DF} R^{\max}_{DF}$ where $R^{\max}_{DF}$ is the DF rate achieved on the best channel and $p^{\max}_{DF}$ is its probability; (ii) Send information via the DF scheme at the rate of the worst (second) channel and hence both users can decode the information at rate $R^{\min}_{DF}$. Finally the next expected rate is achievable by investing on only one coding scheme   
$$
R_{av}^{DF}=\max\big\{p^{\max}_{DF} R^{\max}_{DF},R^{\min}_{DF}\big\};
$$ 
(iii) By investing on CF scheme with the same arguments as before  the next expected rate is also achievable 
$$
R_{av}^{CF}=\max\big\{p^{\max}_{CF} R^{\max}_{CF},R^{\min}_{CF}\big\},
$$ 
with definitions of $(R^{\min}_{CF},R^{\max}_{CF},p^{\max}_{CF})$ similar to before.

Fig. \ref{fig:3} shows numerical evaluation of the average rate. All channel noises are set to the unit variance and $P=P_1=P_2=10$. The distance between $X$ and $(Y_1,Y_2)$ is $(3,1)$, while $d_{z_1}=1$, $d_{z_1y_1}=2$, $d_{z_2}=0.9$, $d_{z_2y_2}=0.1$. As one can see in Fig. \ref{fig:3}, the common rate strategy provides a fixed rate all time which is always better than the worst case. However in one corner the full investments on one rate performs better since the high probability of one channel reduces the effect of the other one. 
Based on the proposed coding scheme, i.e. using the private coding and common coding at the same time, one can cover the corner points and always doing better than both full investments strategies. It is worth to note that in this corner area, only private information of one channel is needed. 

Fig. \ref{fig:4} shows numerical evaluation of $R_0$ for the common-rate case without any probabilistic model. All channel noises are set to the unit variance and $P=P_1=P_2=10$. The distances are set as $d_{y_1}=d_{y_2}=1$ is 1, while $d_{z_1}=d_1$, $d_{z_1y_1}=1-d_1$, $d_{z_2}=0.7$, $d_{z_2y_2}=0.3$. The position of the relay 2 is assumed to be fixed but the relay 1 moves with $d_1\in\left[-1,1\right]$. This setting serves to compare the performances of our coding schemes regarding the position of the relay. It can be seen that one can achieves the minimum between the two possible CF and DF rates. These rates are also compared with a naive time-sharing strategy which consists in using DF scheme $\tau\%$ of the time and CF scheme  $(1-\tau)\%$ of the time\footnote{One should not confuse time-sharing in compound settings with conventional time-sharing which yields convex combination of rates.}. This assumes that the sender uses time-sharing without knowing which channel is present which yields the achievable rate 
$$
R_{TS}=\underset{0\leq\tau\leq1}{\max} \min\{\tau R_{DF},(1-\tau)R_{CF}\}.
$$ 
Notice that with the proposed coding scheme significant gains can be achieved when the relay is close to the source (i.e. DF scheme is more suitable), comparing to the worst case.  


\section{Summary and Discussions}
The simultaneous relay channel with two possible channel outcomes was investigated. The focus was on scenarios where each of the channel outcomes requires a different cooperative strategy. This problem is identified as equivalent to the broadcast relay channel (BRC) where an encoder broadcasts information to two destinations (or channel outcomes) aided by the help of two relays. The coding scheme introduced here enables the source to be oblivious to the relaying strategy. Furthemore it is shown that block Markov coding can be used to simultaneous relaying with DF and CF schemes. Achievable rates were derived and an upper bound on the common rate (or compound case) was also obtained. An application example to the composite Gaussian BRC where the source is unaware of the position of relay, i.e., if relay is collocated to the source or to the destination, was considered. It was shown that significant improvements can be made by using the proposed coding approach comparing to conventional coding strategies. 

As future work, it would be of interest to extend the results in the present work to the setting investigated in \cite{Katz2006} where the source would be oblivious not only to the relying strategy but also to the presence of the relay. 


\section*{Acknowledgment}
 This research is supported by the \emph{Institut Carnot C3S}.  
\bibliographystyle{IEEEtran}
\bibliography{biblio}

\end{document}